\documentstyle[preprint,aps,epsf]{revtex}
\begin{document}
\draft         
\preprint{\font\fortssbx=cmssbx10 scaled \magstep2
\hfill$\vcenter{\hbox{\bf CERN--TH/96--185}
		\hbox{\bf IFT--P.023/96}
}$}
\title{Compositeness Effects in the Anomalous Weak--Magnetic Moment of
Leptons}
\author{M.\ C.\ Gonzalez--Garcia}
\address{Theory Division, CERN,
CH-1211 Geneva 23, Switzerland.}
\author{ S.\ F.\ Novaes }
\address{Instituto de F\'{\i}sica Te\'orica, 
Universidade  Estadual Paulista, \\  
Rua Pamplona 145, CEP 01405-900 S\~ao Paulo, Brazil.}
%
\maketitle
\widetext
\begin{abstract}
We investigate the effects induced by excited leptons, at the
one--loop level, in the anomalous magnetic and weak--magnetic
form factors of the leptons. Using a general effective Lagrangian
approach to describe the couplings of the excited leptons, we
compute their contributions to the weak--magnetic moment of the
$\tau$ lepton, which can be measured on the $Z$ peak, and we
compare it with the contributions to $g_\mu - 2$, measured at low
energies.
\end{abstract}
\noindent
{\bf CERN--TH/96--185} \\
\noindent
{\bf July 1996}\\
\newpage


The standard model of electroweak interactions (SM), in spite of
its remarkable agreement with the present experimental data at
the $Z$ pole \cite{lep:conf}, leaves some important
questions unanswered. In particular, the reason why fermion generations
repeat and the understanding of the complex pattern of quark and
lepton masses are not furnished by the model. With the
proliferation of fermion flavours, it is natural to ask whether
these particles are truly elementary states. The idea of
composite models assumes the existence of an underlying 
structure, characterized by a mass scale $\Lambda$, with the
fermions sharing  some of the constituents \cite{comp}.  As a
consequence, excited states of each known lepton  should show up
at some energy scale, and the SM should be seen as the
low--energy limit of a more fundamental theory.  

We still do not have a satisfactory model, able to
reproduce the whole particle spectrum. Due to the lack of a
predictive theory, we should rely on a model--independent
approach to explore the possible effects of compositeness,
employing effective Lagrangian techniques to describe the
couplings of these excited states.

Several experimental collaborations have been searching for
excited lepton states \cite{rev:lep,hera}. Their analyses are
based on an effective  $SU(2) \times U(1)$ invariant Lagrangian,
proposed some years ago by Hagiwara {\it et al.\/} \cite{hag}.
Also a series of phenomenological studies of excited fermions
have been carried out in 
electron--positron  \cite{hag,ele:pos,e:gam,kuhn,pp,lep:hera}, 
hadronic \cite{kuhn,pp}, and
electron--proton  \cite{hag,lep:hera} collisions. 

On the other hand, an important source of indirect information
about new particles and interactions is the precise measurement
of the electroweak parameters. Virtual effects of these new
states can alter the SM predictions for some of these parameters, 
and the comparison with the experimental data can impose bounds
on their masses and couplings. Bounds have been derived from the
contribution to the anomalous magnetic moment of leptons
\cite{g-2} and to the $Z$ observables at LEP \cite{exciZ}. 

In this work we use a general effective Lagrangian approach to
investigate the effects induced by excited leptons, at the
one--loop level, in the anomalous  weak--magnetic form factors of
the leptons, at an arbitrary energy scale. In particular, we 
study the contribution  to the weak--magnetic moment of the
$\tau$ lepton that can be measured on the $Z$ peak \cite{berna1}.
Our results show that for universal couplings the existing limits
from $g_\mu-2$ strongly constrain the possibility of observing
this effect on the anomalous weak--magnetic moment of the $\tau$
lepton at LEP, given the expected experimental sensitivity
\cite{berna1}.

We consider excited fermionic states with  spin and isospin
$\frac{1}{2}$, and we assume that the excited fermions acquire
their masses before the $SU(2) \times U(1)$ breaking, so that
both left--handed and right--handed states belong to weak
isodoublets. The most general dimension--six effective Lagrangian
\cite{hag,lep:hera} that describes the coupling of the
excited--usual fermions, which is  $SU(2) \times U(1)$ invariant
and CP-conserving, can be written as
\begin{equation}
{\cal L}_{Ff} = - \sum_{V=\gamma,Z,W} 
C_{VFf} \bar{F} \sigma^{\mu\nu} (1 - \gamma_5) f \partial_\mu V_\nu 
- i \sum_{V=\gamma,Z} D_{VFf} \bar{F} \sigma^{\mu\nu} (1 - \gamma_5) f
W_\mu V_\nu + \; \text{h.c.} \; ,
\label{l:eu}
\end{equation}
where $F = N, E$ represent the excited states, and $f = \nu, e$,
the usual light fermions of the first generation. A pure
left--handed structure is assumed for these couplings in order to
comply with  the strong bounds coming from the measurement of the
anomalous magnetic  moment of leptons \cite{g-2}.  The coupling
constants $C_{VFf}$ are given by
\begin{equation}
\begin{array}{ll}
C_{\gamma E e} =  - \frac{\displaystyle e}{\displaystyle 4 \Lambda} 
(f_2 + f_1) &
\;\; , \;\;\;\;
C_{\gamma N \nu} =  
\frac{\displaystyle e}{\displaystyle 4 \Lambda} (f_2 - f_1)\\
C_{Z E e} =  
- \frac{\displaystyle e}{\displaystyle 4 \Lambda} 
(f_2 \cot\theta_W - f_1 \tan\theta_W) &
\;\; , \;\;\;\;
C_{Z N \nu} =  
\frac{\displaystyle e}{\displaystyle 4 \Lambda} 
(f_2 \cot\theta_W + f_1 \tan\theta_W)\\
C_{W E \nu} =   C_{W N e} = 
\frac{\displaystyle e}{\displaystyle 2\sqrt{2}\sin\theta_W\Lambda}f_2 \; , & 
\end{array}
\label{CV}
\end{equation}
where $\theta_W$ is the weak mixing angle, $f_2$ and $f_1$ are
weight factors associated to the $SU(2)$ and $U(1)$ coupling
constants, and $\Lambda$ is the compositeness scale. The quartic
interaction coupling constant, $D_{VFf}$, is given by,
\begin{equation}
\begin{array}{l}
D_{\gamma E \nu} = - D_{\gamma N e} =  
- \frac{\displaystyle e^2 \sqrt{2}}
{\displaystyle 4 \sin\theta_W \Lambda} f_2\\
D_{Z E \nu} = - D_{Z N e} = 
- \frac{\displaystyle e^2 \sqrt{2} \cos\theta_W}
{\displaystyle 4 \sin^2\theta_W \Lambda} f_2\; .
\end{array}
\label{DV}
\end{equation}

The coupling of gauge bosons to excited leptons can also be
described by the $SU(2) \times U(1)$ invariant and CP-conserving 
effective Lagrangian,
\begin{equation}
{\cal L}_{FF} = - \sum_{V=\gamma,Z,W} \bar{F} 
(A_{VFF} \gamma^\mu V_\mu + K_{VFF} \sigma^{\mu\nu} \partial_\mu V_\nu) F
\; ,
\label{l:ee}
\end{equation}
where $A_{VFF}$ is given by
\begin{equation}
\begin{array}{ll}
A_{\gamma EE} =  - e &
\;\; , \;\;\;\;
A_{\gamma NN} =   0 \\
A_{ZEE} = e \frac{\displaystyle (2 \sin^2\theta_W - 1)}
{\displaystyle 2  \sin\theta_W  \cos\theta_W} &
\;\; , \;\;\;\; 
A_{ZNN} =  \frac{\displaystyle e}{\displaystyle 2\sin\theta_W \cos\theta_W}
\\
A_{WEN} = \frac{\displaystyle e}{\displaystyle \sqrt{2} \sin\theta_W} &
\end{array}
\label{AV}
\end{equation}
and $K_{VFF}$ is given by
\begin{equation}
\begin{array}{ll}
K_{\gamma EE} =
- \frac{\displaystyle e}{\displaystyle 4 \Lambda} (\kappa_2 + \kappa_1) & 
\;\; , \;\;\;\;
K_{\gamma NN} =  
\frac{\displaystyle e}{\displaystyle 4 \Lambda} (\kappa_2 - \kappa_1)\\
K_{Z EE} = 
- \frac{\displaystyle e}{\displaystyle 4 \Lambda}
(\kappa_2\cot\theta_W -\kappa_1\tan\theta_W)  &
\;\; , \;\;\;\;
K_{Z NN} =  
\frac{\displaystyle e}{\displaystyle 4 \Lambda} 
(\kappa_2 \cot\theta_W + \kappa_1\tan\theta_W)
\\ 
K_{WEN} =  
\frac{\displaystyle e}{\displaystyle 2 \Lambda} 
\frac{\displaystyle \kappa_2}{\displaystyle \sqrt{2} \sin\theta_W}.
\end{array}
\label{KV}
\end{equation}

The matrix element of a boson ($V_1$) current has the general form: 
\begin{equation}
J^\mu  = e\; \bar u_f (p_1) \;  
\left\{ \frac{1}{2 \sin\theta_W \cos\theta_W} \gamma^\mu 
\left[ F_V(q^2) - F_A(q^2) \gamma^5 \right]
+ \frac{i}{2 m_f} a_f^{V_1}(q^2) \sigma^{\mu\nu} q_\nu \right\} 
v_f (p_2)  
\label{form:nc}
\end{equation}
where $V_1 = \gamma$, or $Z$, and $q = p_1 + p_2$. The terms
$F_V$ and $F_A$ are present at tree level in the SM, {\it e.g.\/}
for the $Z$ boson, $F_V^{\text{tree}} = - T_3^f + 2 Q_f \sin^2
\theta_W$, and $F_A^{\text{tree}} = - T_3^f$. The contribution of
the excited leptons to these form factors at the one--loop level has
been evaluated in Ref. \cite{exciZ}. The anomalous weak--magnetic
form factor, $a_f^{V_1}$, is generated only at one--loop in the
SM as well as in the models with excited fermions. In the latter
case, there are twelve one--loop Feynman diagrams involving
excited fermions that contribute to the anomalous
electroweak--magnetic moment of leptons, which are shown in Fig.\
\ref{fig:1}. For each of these contributions, we define the
amplitudes  $S^{V_2}_i (q^2, M^2, M_{V_2}^2)$, $i = 1, \cdots ,
12$, where $V_2$ is the virtual vector boson with mass $M_{V_2}$
running in the loops, and we can write the excited lepton
contribution to $a_f^{V_1}$ as
\begin{equation}
a_f^{V_1}(q^2)=  i \frac{m_f^2 }{e} \; S_{V_1\rightarrow \bar{f} f} (q^2)\; ,
\label{mat}
\end{equation}
with
\begin{equation}
\begin{array}{ll}
S_{V_1\to f^+ f^-}(q^2) = & 
S_1^\gamma (q^2, M^2, 0) + S_1^Z (q^2, M^2, M_Z^2) + S_1^W (q^2, M^2, M_W^2) 
\\
& +
S_{2+3}^\gamma (q^2, M^2, 0) + S_{2+3}^Z (q^2, M^2, M_Z^2) 
+ S_{2+3}^W (q^2, M^2, M_W^2)  
\\
& + 
S_4^W (q^2, M^2, M_W^2) \\
& +
S_{7+8}^\gamma (q^2, M^2, 0) + S_{7+8}^Z (q^2, M^2, M_Z^2) + 
S_{7+8}^W (q^2, M^2, M_W^2) 
\\
& +
S_{9+10}^\gamma (q^2, M^2, 0) + S_{9+10}^Z (q^2, M^2, M_Z^2) + 
S_{9+10}^W (q^2, M^2, M_W^2) 
\\
& + S_{11+12}^W (q^2, M^2, M_W^2)  \; .
\end{array}
\label{g:ee}
\end{equation}
We have neglected the fermion masses in the evaluation of the
integrals $S^{V_2}_i$ ({\it i.e.\/}  $m_f^2 \ll M^2, M_V^2$),
and in this limit $S^{V_2}_{5+6} (q^2,M^2, M_V^2) = 0$.

The loop contributions of the excited leptons were evaluated in
$D = 4 - 2 \epsilon$ dimensions using the dimension
regularization method \cite{reg:dim}, which is a gauge--invariant
regularization procedure, and we adopted the unitary gauge to
perform the calculations. The results in $D$ dimensions were
obtained with the aid of the Mathematica package FeynCalc
\cite{feyn}, and the poles at $D=4$ ($\epsilon=0$) and $D=2$
($\epsilon= 1$) were identified with the logarithmic and
quadratic dependence on the scale $\Lambda$ \cite{zep}. 

Our results for $S^{V_2}_{i} (q^2, M^2, M_{V_2}^2)$ are rather 
lengthy. We show here only approximate expressions, which are valid in
the limit $q^2, M_V^2 \ll M^2$, at first order in  $R_Q= q^2/M^2$
and $R_V=M_V^2/M^2$:
\begin{equation}
\begin{array}{ll}
S^{V_2}_1
\simeq & 
\frac{\displaystyle  i}{\displaystyle 144 \pi^2 }\, C^2_{V_2Ff} 
\,
\Biggl[ 6 A_{V_1FF}(20 + 9 R_V) + 36 K_{V_1FF}\,M (3 + R_V) \\
 & - R_Q ( 15 A_{V_1FF}  + 32 K_{V_1FF}\,M ) 
- 72 \left(A_{V_1FF}\, +\,K_{V_1FF}\,M\right) 
\log \frac{\displaystyle \Lambda^2}{\displaystyle M^2} \Biggr]
\\
S^{V_2}_{2+3}
\simeq & 
\frac{\displaystyle i}{\displaystyle 24 \pi^2 }\, C_{V_2Ff}C_{V_1Ff} 
\Biggl\{
g_{V_2}^v \Bigl[3 + 6 R_V + 12 R_V \log R_V  +
4 R_Q \left( 1 + 2 \log R_V \right ) 
+6 \log \frac{\displaystyle \Lambda^2}{\displaystyle M^2} \Bigr]
\\ 
& +
g_{V_2}^a \Bigl( 39 + 6 R_V + 12 R_V \log R_V -16 R_Q  \log R_V 
+ 6 \log \frac{\displaystyle \Lambda^2}{\displaystyle M^2} \Bigr)
\Biggr\}
\\
S^{V_2}_{4}
\simeq & 
\frac{\displaystyle i}{\displaystyle 144 \pi^2 }\, C^2_{V_2Ff} g_{V_1 WW}
\Biggl( 79  - 9 R_Q  + 120 R_V - 42 
\log \frac{\displaystyle \Lambda^2}{\displaystyle M^2}
\Biggr)
\\
S^{V_2}_{7+8}
\simeq & 
\frac{\displaystyle i}{\displaystyle 2 \pi^2 }\, C_{V_1Ff}C_{V_2Ff} 
\left(g_{V_2}^a+g_{V_2}^v \right)
\Biggl(2 R_V +3 R_V \log R_V -3 R_V  
\log \frac{\displaystyle \Lambda^2}{\displaystyle M^2} 
+ \frac{\displaystyle \Lambda^2}{\displaystyle M^2} \Biggr)
\\
S^{V_2}_{9+10}
\simeq & 
\frac{\displaystyle i}{\displaystyle 8 \pi^2 }\, C_{V_1Ff}C_{V_2Ff} 
\Biggl\{ A_{V_1FF}(15 + 14 R_V) + K_{V_1FF}\,M (22 + 21 R_V)  \\
& - 6 \Bigl[A_{V_1FF}(3 + 2 R_V) +K_{V_1FF}\,M (4 + 3 R_V) \Bigr]
\log \frac{\displaystyle \Lambda^2}{\displaystyle M^2} 
+ 4 \frac{\displaystyle \Lambda^2}{\displaystyle M^2} 
\left(A_{V_2FF}\, +\,K_{V_2FF}\,M\right) \Biggr\}
\\
S^{V_2}_{11+12}
\simeq & 
\frac{\displaystyle i }{\displaystyle 144 \pi^2}
\,C_{V_2Ff}\,D_{V_1Ff}\,
\Biggl( 1 + 12 R_V - 6  
\log \frac{\displaystyle \Lambda^2}{\displaystyle M^2} \Biggr)
\end{array}
\label{S}
\end{equation}
where $g^v_{V_2}$ and $g^a_{V_2}$ are the vector and axial coupling of the
vector bosons to the usual fermions: 
$g^v_{\gamma} = - e$, $g^a_{\gamma} = 0$;  
$g^v_{W} =  g^a_{W} =  g/(2 \sqrt{2})$; 
for $f = \nu$, $g^v_{Z} =  g^a_{Z} = g/(4 \cos\theta_W)$;
for  $f = e$, $g^v_{Z} = g (4 \sin^2\theta_W - 1)/( 4
\cos\theta_W)$,  and  $g^a_{Z} = - g/(4 \cos\theta_W)$. 
The  coupling $g_{V_1WW}$ refers to the triple vector
boson vertex: $g_{\gamma WW} =
g \sin\theta_W$, and $g_{ZWW} = g \cos\theta_W$.

We should notice that $S_2^\gamma$ is infrared--divergent for $q^2
\neq 0$. However, this divergence cancels against the one coming
from real photon emission, and the final result is therefore
infrared--finite. In the appendix we present the details of this
cancellation. 

We give here the approximate final results for the  anomalous
magnetic ($a_f^{\gamma}$) and weak--magnetic ($a_f^{Z}$) moments,
assuming $M^2 = \Lambda^2 \gg M_{W,Z}^2$, and $f_1=f_2=f$ and
$k_1=k_2=k$: 
\begin{equation}
\begin{array}{l}
a_f^{\gamma}= \frac{\displaystyle \alpha}{\displaystyle 48 \pi} 
\frac{\displaystyle f^2 \,m_f^2}{\displaystyle M^2} 
\left[ \frac{\displaystyle 37 + 74 \cos^2\theta_W +(24 +39 \cos^2\theta_W) k}
{\displaystyle \sin^2\theta_W \cos^2\theta_W}  \right] \\
a_f^{Z}= - \frac{\displaystyle \alpha}{\displaystyle 96 \pi} 
\frac{\displaystyle f^2 \, m_f^2}{\displaystyle M^2} 
\left[ \frac{\displaystyle 37 + 2 \cos^2\theta_W (27  - 74 \cos^2\theta_W) + 
6 (4 - 13 \cos^4\theta_W) k}
{\displaystyle \sin^3\theta_W \cos^3\theta_W}  \right] \; .
\end{array}
\label{moments}
\end{equation} 
Our results for the anomalous magnetic moment $a_f^{\gamma}$ are
in agreement with those of Ref. \cite{g-2}, for $k=0$. 

We now turn to the attainable values for the weak--magnetic
moment of the $\tau$ lepton at LEP energies. In Ref.\
\cite{berna1},  Bernabeu {\sl et al.\/} compute the SM
contribution to this observable and discuss the attainable
sensitivity at LEP. They claim that it  can be measured through
the analysis of the angular asymmetry of the semileptonic $\tau$
decay products, which carries information about the
weak--magnetic moment of the parent lepton. They assume that the
$\tau$ direction is fully reconstructed and they deduce a
sensitivity of the order of $|a^Z_\tau (M_Z^2)| \lesssim
10^{-4}$.

\begin{figure}
\protect
\epsfysize=19cm
\begin{center}
\leavevmode \epsfbox{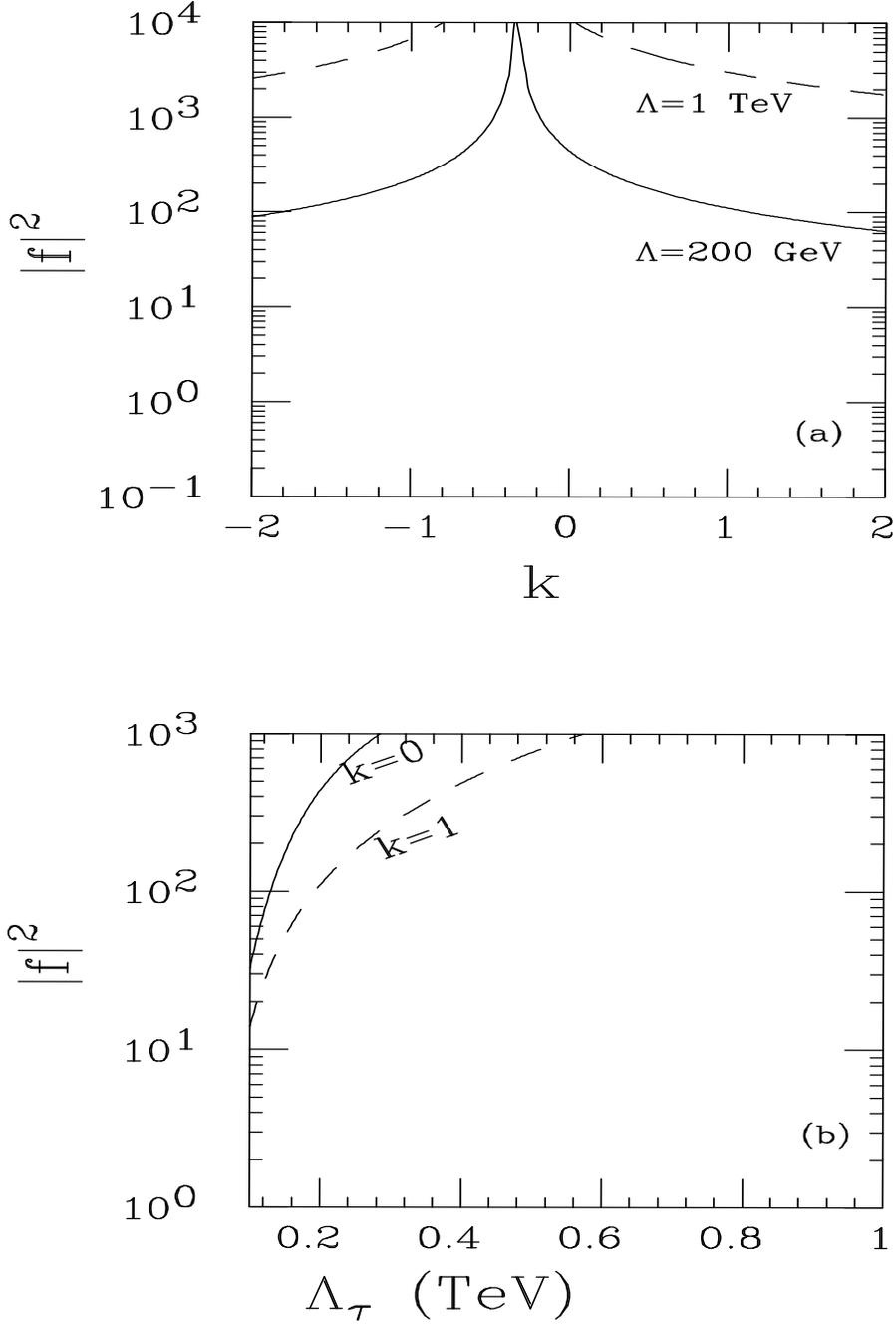}
\end{center}
\caption{{\bf (a)}  Accessible region of parameters for 
$|a_\tau^Z(M_Z^2)|\geq 10^{-4}$ (above the curves) in the $|f|^2$ versus 
$k$ plane for fixed values of $M=\Lambda$. 
{\bf (b)}  Accessible region (above the curves) in the $|f|^2$ versus
$M=\Lambda$ plane for fixed values of $k$.}
\label{fig:2}
\end{figure}

In Fig. \ref{fig:2}, we show the accessible region in the
parameter space  $f$, $k$ and $\Lambda$. For the sake of
simplicity we have assumed that $\Lambda=M$. As seen in this
figure only models with strong coupling, {\it i.e.\/} $f \simeq
\sqrt{4 \pi}/e$, and compositeness scale $\Lambda\lesssim 200$ GeV
could lead to a value for the anomalous weak--magnetic moment of
the $\tau$ large enough to be observed at LEP. 

If we assume that the couplings to the excited fermions are
universal, {\it i.e.\/} if $f_i$, $k_i$, $M$ and $\Lambda$ are the
same for the three generations, the attainable value for the
$\tau$ lepton weak--magnetic moment is already constrained by the
existing limits from the anomalous electromagnetic moment of the
muon measured at low energies.  Nowadays, the most precise
determination of the anomalous magnetic moment of the muon
$a^\gamma_\mu \equiv (g_\mu - 2)/2$ comes from a CERN experiment
\cite{g-2exp}:
\begin{equation}
a^\gamma_\mu= 11 \, 659 \, 230 \, (84) \times 10^{-10} .
\end{equation}
This result should be compared with the existing theoretical
calculations of the QED \cite{g-2QED}, electroweak \cite{g-2EW}
and hadronic \cite{g-2had} contributions, which are known with high
precision.  The main theoretical uncertainty comes
from the hadronic contributions which is of the order of $20
\times 10^{-10}$. Therefore the present limit on the
non--standard contributions to the anomalous magnetic moment of
the muon is 
\begin{equation}
|\delta \, a^\gamma_\mu| \lesssim 8 \times 10^{-9}\; .
\label{g-2lim}
\end{equation}   
The proposed AGS experiment at the Brookhaven National Laboratory 
\cite{futu} will be able to  measure the anomalous magnetic
moment of the muon with an accuracy of about  $\pm 4 \times
10^{-10}$.

\begin{figure}
\protect
\epsfysize=19cm
\begin{center}
\leavevmode \epsfbox{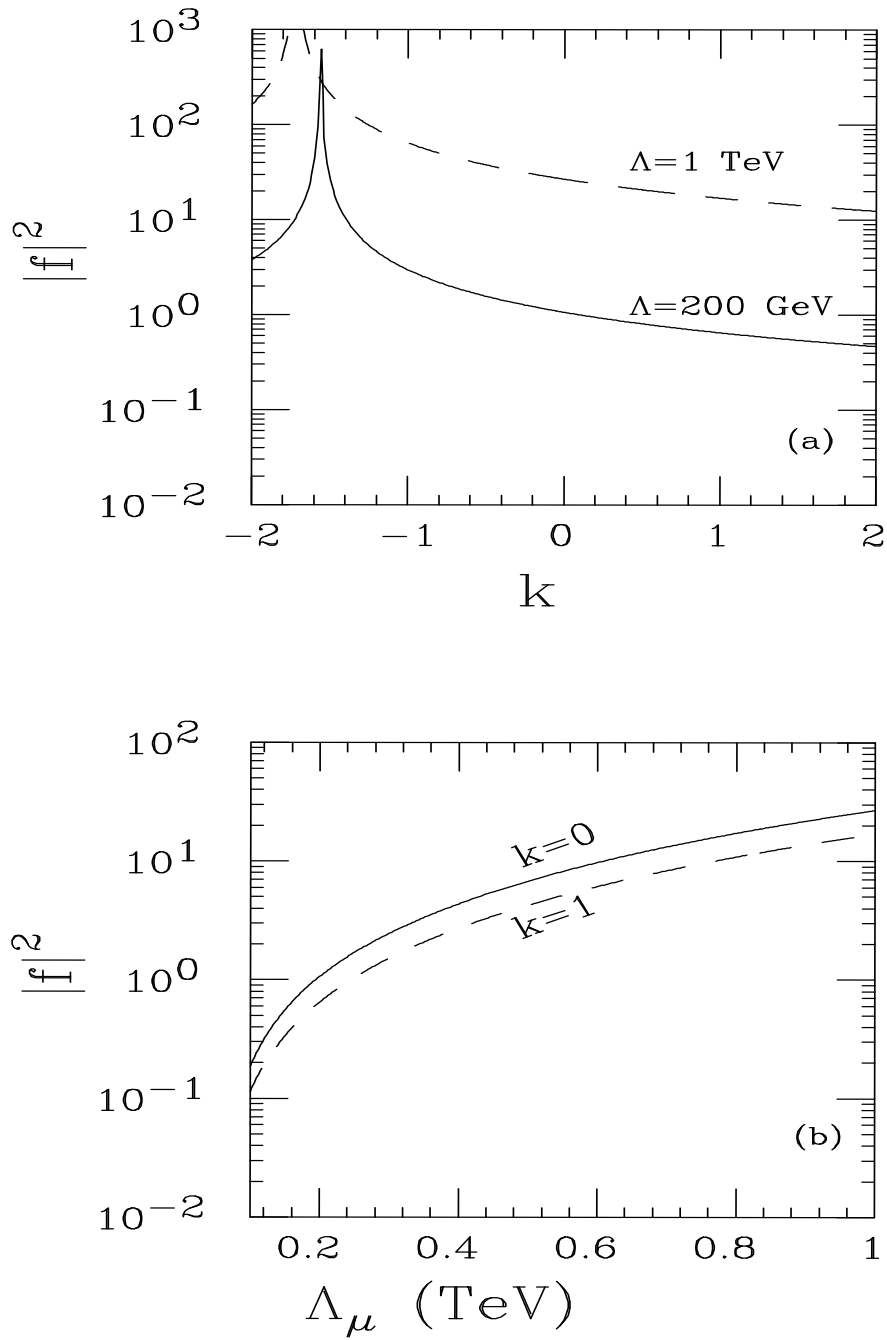}
\end{center}
\caption{{\bf (a)}  Excluded  region of parameters from 
$|a_\mu^\gamma|\leq 8\times 10^{-9}$ (above the curves) 
in the $|f|^2$ versus  $k$ plane for fixed values of $M=\Lambda$. 
{\bf (b)}  Corresponding limits in the  $|f|^2$ versus
$M=\Lambda$ plane for fixed values of $k$.}
\label{fig:3}
\end{figure}

In Fig.\ \ref{fig:3} we show the present limits from Eq.\
(\ref{g-2lim})  in the parameter space  $f$, $k$ and $\Lambda$. Our
results for $k=0$ are in agreement with those from Ref.
\cite{g-2}. However, the presence of an anomalous magnetic moment
term  at tree level in the coupling between a pair of excited
fermions (see Eq.\ (\ref{l:ee})) alters the attainable bounds on
$f$. As seen in Fig. \ref{fig:3}a,  there is a value  $k = k_0
\simeq -1.56$ ($-1.72$), for $\Lambda = 0.2$ ($1.0$) TeV,  for
which the limit on the coupling strength $f$ becomes very weak.
This comes as a consequence of the cancellation of the leading
terms in Eq.\ (\ref{moments}). The exact dependence of $k_0$ on
$\Lambda$ is due to higher--order terms, which are not displayed in
Eq.\ (\ref{moments}).

\begin{figure}
\protect
\epsfxsize=15cm
\begin{center}
\leavevmode \epsfbox{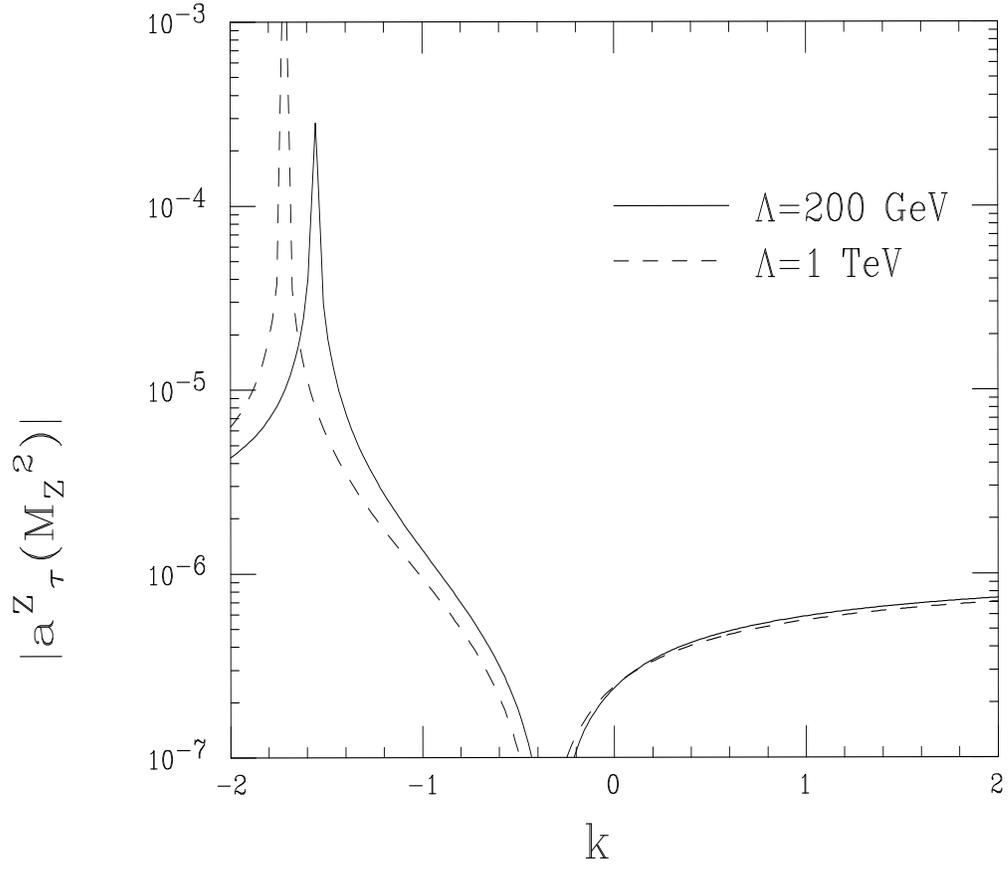}
\end{center}
\caption{Attainable values of $|a_\tau^Z(M_Z^2)|$ for universal
excited lepton couplings after imposing the constraints from
$g_\mu-2$.}
\label{fig:4}
\end{figure}

Taking these results into account, we plot in Fig. \ref{fig:4}
the attainable values for the $\tau$ anomalous weak--magnetic
moment, assuming universal couplings, after imposing  the
constraints from $g_\mu-2$ measurements. We can see that only for
a narrow band of $k$ values around $k_0$ can $|a^Z_\tau (M_Z^2)|$
be large enough to be observed at LEP. 

Summarizing, we have investigated the effects induced by excited
leptons at the one--loop  level in the anomalous magnetic and
weak--magnetic form factors of the  leptons at an arbitrary scale.
Using a general effective Lagrangian approach to describe the
couplings of the excited leptons, we have computed their
contributions to the  weak--magnetic moment of the $\tau$,  which
can be measured on the $Z$ peak, and we compare it with the
contributions to $g_\mu-2$ measured at low energies. Our results
show that although for universal couplings, the existing limits
from  $g_\mu-2$ strongly constrain the possibility of observing
the anomalous weak--magnetic moment of the $\tau$ lepton at LEP,
there is a very narrow region of parameters in which the
observation is still possible. One must notice, however, that in
this region the model becomes strongly coupled.   

\newpage
\appendix
\section{Cancellation of the Infrared Divergences}

\begin{figure}
\protect
\epsfxsize=14cm
\begin{center}
\leavevmode \epsfbox{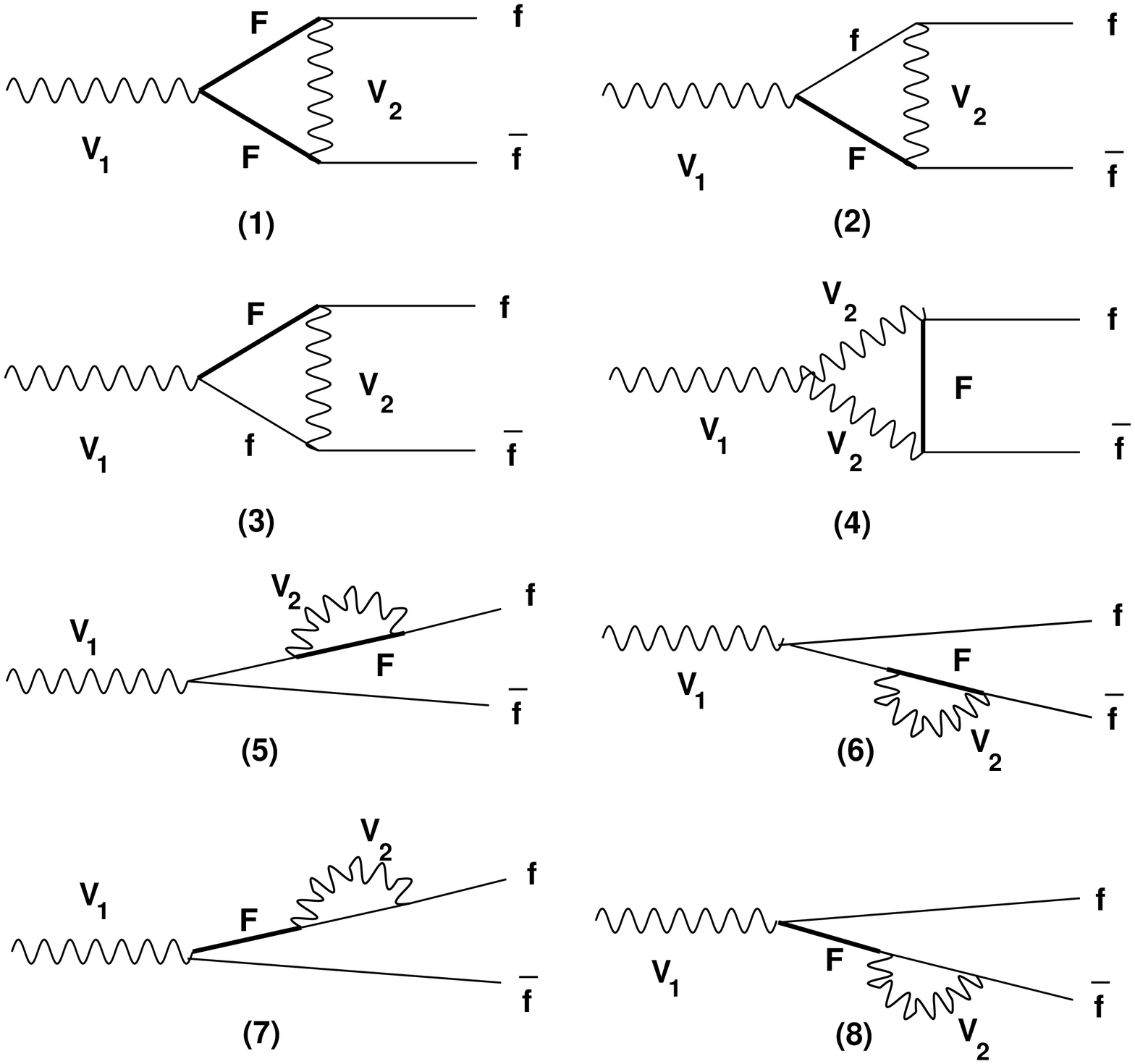}
\end{center}
\epsfxsize=14cm
\begin{center}
\leavevmode \epsfbox{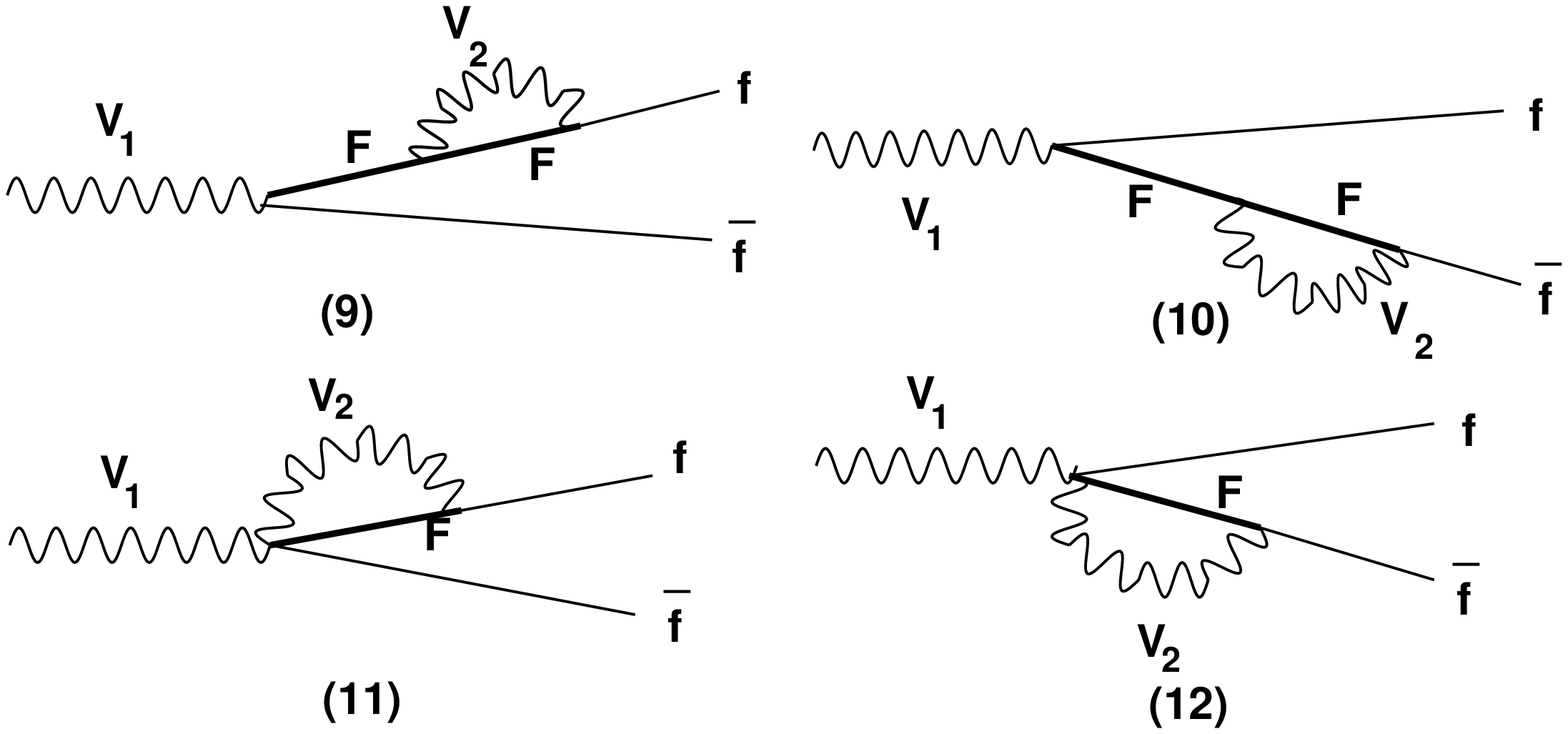}
\end{center}
\caption{The contribution of the excited leptons to the anomalous
electroweak magnetic moments.}
\label{fig:1}
\end{figure}

The one--loop contribution of excited fermions to the amplitude
$Z\rightarrow f\bar f$ presents an infrared--divergent piece due
to diagrams 2 and 3 of Fig.\ \ref{fig:1}, for  $V_2 = \gamma$. At
order $m_f^2$ the exact expression for the infrared contribution
for any $q^2$ is
\begin{eqnarray*} 
M^{\text{IR}}_{2+3} &=& -\bar u (q_1) \sigma^{\mu\nu} q^\nu 
\epsilon^\mu_{V_1}(q) v(q_2) \\
& \times & \left\{ \frac{m_f}{4 \pi^2 q^4}
C_{V_2Ff} C_{V_1Ff} g_{\gamma}^v \left[ 2 M^2 q^2 + q^4 +2 M^4
\log \left( 1-\frac{q^2}{M^2} \right) \right] \right\} \; ,
\log\frac{\lambda^2}{M^2}
\end{eqnarray*}
where by $\lambda$ we have denoted the photon mass. 
At order $\alpha^2$, this amplitude contributes to the
$Z\rightarrow f\bar f$ decay width via the interference with the
tree--level amplitude
\[
M_0=-i\bar u (q_1) \gamma^{\mu} \epsilon^\mu_{V_1}(q) 
(g_{V_1}^v-g_{V_1}^a \gamma^5) v(q_2)\; ,
\]
yielding
\begin{equation}
\Gamma_{1-\text{loop}}^{\text{IR}} = -\frac{m_f^2}{24 \pi^3 q^3} 
C_{V_2Ff}C_{V_1Ff} g_{\gamma}^v g_{V_1}^v \left[ 2 M^2 q^2 + q^4 +2 M^4
\log\left( 1-\frac{q^2}{M^2} \right) \right]\log\frac{\lambda^2}{M^2}\; .
\label{G1loop}
\end{equation}

\begin{figure}
\protect
\epsfxsize=15cm
\begin{center}
\leavevmode \epsfbox{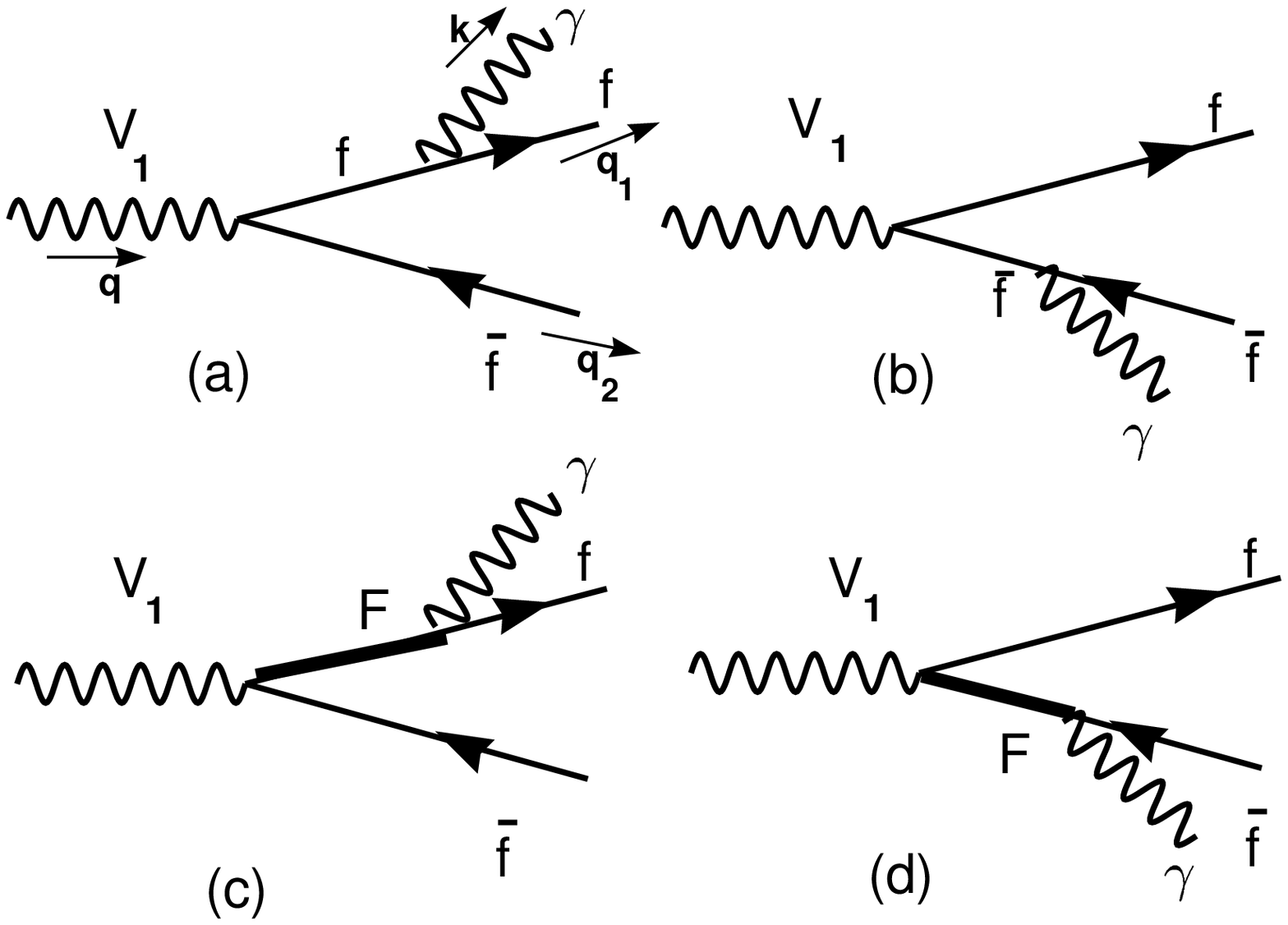}
\end{center}
\caption{Diagrams for photon bremsstrahlung.}
\label{fig:5}
\end{figure}

This divergence is cancelled against the one coming from the 
photon  bremsstrahlung processes presented in Fig.\ \ref{fig:5}.
At order $\alpha^2$, the contribution from these diagrams ($M_{a
\cdots d}$) to the decay width $Z(q) \to f (q_1) + \bar{f}(q_2) +
\gamma (k) $ is given by:
\begin{eqnarray*}
\Gamma_{\text{bremss}} &=& \frac{1}{2 q (2\pi)^5}
\int 
\frac{d^3 q_1}{2 E_1}
\frac{d^3 q_2}{2 E_2}
\frac{d^3 k}{2 k_0} \delta^4(q - q_1 - q_2 - k) 
\frac{2}{3}
\text{Re} \left[(M_a+M_b)^\dagger(M_c+M_d)\right] \\
& =& \frac{1}{32 q^3 (2 \pi)^3}
\int_{s_{2\text{min}}}^{s_{2\text{max}}} ds_2 
\int_{s_{1+}}^{s_{1-}} ds_1 F(s_1, s_2) \; ,
\end{eqnarray*}
where $s_{2\text{min}}=(m_f+\lambda)^2$ and
$s_{2\text{max}}=(q-m_f)^2$, and  
\[
s_{1\pm}= m_f^2+\lambda^2 
-\frac{\displaystyle 1}{\displaystyle 2 s_2}
\left[(s_2-q^2+m_f^2)(s_2+\lambda^2 -m_f^2)\pm
f^{1/2}(s_2,q^2,m_f^2) f^{1/2}( s_2,\lambda^2,m_f^2)\right]
\] 
with the function $f(a,b,c) = (a-b-c)^2-4 b c$.  

At leading order in $m_f^2$, the infrared--divergent piece of
$F(s_1, s_2)$  originates from 
\[
F^{\text{IR}}(s_1,s_2)=-\frac{32 \, m_f^2}{3} C_{V_2Ff}C_{V_1Ff} g_{\gamma}^v 
g_{V_1}^v \Biggl[ \frac{(s_1 - m_f^2)^2}{(s_1 - M^2)(s_2 - m_f^2)}
+ \frac{(s_2-m_f^2)^2}{(s_2 - M^2)(s_1 - m_f^2)}
\Biggr] \; ,
\]
and we finally get the infrared--divergent contribution from
bremsstrahlung as
\[
\Gamma_{\text{bremss}}^{\text{IR}}=\frac{m_f^2}{24 \pi^3 q^3} 
C_{V_2Ff}C_{V_1Ff} g_{\gamma}^v g_{V_1}^v \Biggl[2 M^2 q^2 + q^4 +2 M^4
\log\left(1-\frac{q^2}{M^2}\right)\Biggr]\log\frac{\lambda^2}{M^2}
=-\Gamma_{1-\text{loop}}^{\text{IR}}
\]
which exactly cancels the contribution from Eq. (\ref{G1loop}).




\begin{references}

\bibitem{lep:conf} The LEP Collaborations ALEPH, DELPHI, L3,
OPAL, and the LEP Electroweak Working Group, contributions to the
1995 Europhysics Conference  on High Energy Physics
(EPS--HEP), Brussels, Belgium, and to the 17th International
Symposium on Lepton--Photon Interactions,  Beijing, China,
Report No. CERN--PPE/95--172 (1995).

\bibitem{comp} For a review, see for instance:   H.\ Harari,
Phys.\ Reports {\bf 104} (1984)  159;  H.\ Terazawa, Proceedings
of the XXII International Conference on High Energy Physics,
Leipzig, 1984, edited by A.\ Meyer and E.\ Wieczorek, p.\ 63; W.\
Buchm\"uller, Acta Phys.\ Austriaca, Suppl.\ XXVII (1985) 517;
M.\ E.\ Peskin, Proceedings of the 1985  International Symposium
on Lepton and Photon Interactions at High Energies, Kyoto, 1985, p.\ 714, 
eds. M.\ Konuma and K.\ Takahashi.

\bibitem{rev:lep} ALEPH Collaboration, D.\ Decamp {\it et al.}, 
Phys.\ Lett.\ {\bf B236} (1990) 501; {\it id}, {\bf B250} (1990) 172;  \\
L3 Collaboration, B.\ Adeva {\it et al.}, Phys.\ Lett.\ 
{\bf B247} (1990) 177; {\it id}, {\bf B250} (1990) 199 and 205; 
{\it id}, {\bf B252} (1990) 525;  
L3 Collaboration, O.\ Adriani  {\it et al.}, Phys.\ Lett.\ 
{\bf B288} (1992) 404; 
L3 Collaboration, M.\ Acciarri  {\it et al.}, Phys.\ Lett.\ 
{\bf B353} (1995) 136, {\it id},  {\bf B370} (1996) 211; \\
OPAL Collaboration, M.\ Z.\ Akrawy {\it et al.}, Phys.\ Lett.\
{\bf B240} (1990) 497; {\it id}, {\bf B241} (1990) 133; 
{\it id}, {\bf B244} (1990) 135; {\it id}, {\bf B257} (1991) 531; \\
DELPHI Collaboration, P.\ Abreu {\it et al.}, Phys.\ Lett.\ 
{\bf B268} (1991) 296; {\it id}, {\bf B327} (1994) 386;
Z.\ Phys.\  {\bf C53} (1992) 41.

\bibitem{hera} H1 Collaboration, I.\ Abt {\it et al.}, Nucl.\
Phys.\ {\bf B396} (1993) 3;  
ZEUS Collaboration, M.\ Derrick {\it et al.}, Phys.\ Lett.\ {\bf
B316} (1993) 207, {\it id}, Z.\ Phys.\  {\bf C65} (1994) 627.

\bibitem{hag} K.\ Hagiwara, S.\ Komamiya and D.\ Zeppenfeld, Z.\
Phys.\ {\bf C29} (1985) 115.

\bibitem{ele:pos} N.\ Cabibbo, L.\ Maiani  and Y.\ Srivastava,
Phys.\ Lett.\ {\bf B139} (1984) 459;  
F.\ A.\ Berends and P.\ H.\ Daverveldt, Nucl.\ Phys.\ {\bf B272} (1986)
131;  
A.\  Feldmaier, H.\  Salecker and F.\  C.\  Simm, Phys.\  Lett.\
{\bf B223} (1989) 234;  
M.\ Martinez, R.\ Miquel and C.\ Mana, Z.\ Phys. {\bf C46} 
(1990) 637;  
F.\ Boudjema and A.\ Djouadi, Phys.\ Lett.\ {\bf B240} 
(1990) 485;  
M. Bardadin--Otwinowska, Z.\ Phys.\ {\bf C55} (1992) 163;  
J.\ C.\ Montero and V.\ Pleitez,  Phys.\ Lett.\ {\bf B321} 
(1994) 267.

\bibitem{e:gam} I.\ F.\ Ginzburg and D.\ Yu.\ Ivanov, Phys.\
Lett.\ {\bf B276} (1992)  214;  
T.\ Kon, I.\ Ito and Y.\ Chikashige,  Phys.\ Lett.\ {\bf B287} (1992)
277;  
E.\ Boos, A.\ Pukhov and A.\ Beliaev, Phys.\ Lett.\ {\bf B296} 
(1992) 452;
O.\ J.\ P.\ \'Eboli, E.\ M.\ Gregores, J.\ C.\ Montero,  S.\ F.\
Novaes and D.\ Spehler, Phys.\ Rev.\ {\bf D53} (1996) 1253.


\bibitem{kuhn} J.\ K\"uhn and P.\ Zerwas, Phys.\ Lett.\ {\bf
B147} (1984) 189.

\bibitem{pp} K.\ Enqvist and J.\ Maalampi,  Phys.\ Lett.\ {\bf
B135} (1984) 329.

\bibitem {lep:hera} F.\ Boudjema, A.\ Djouadi and J.\ L.\ Kneur,
Z.\ Phys. {\bf C57} (1993) 425.

\bibitem{g-2} S.\ J.\ Brodsky and S.\ D.\ Drell,  Phys.\ Rev.\
{\bf D22} (1980) 2236; F.\ M.\  Renard, Phys.\  Lett.\  {\bf
B116} (1982) 264; P.\ Merry, S.\ E.\ Moubarik, M.\ Perrottet and 
F.\ M.\ Renard, Z. Phys. {\bf C 46} (1990) 229;
R.\ Escribano and E.\ Masso, hep--ph/9607218. 

\bibitem{exciZ} M.\ C.\ Gonzalez--Garcia and S.\ F.\ Novaes,
CERN--TH/96--123  and IFT--P.013/96. 

\bibitem{berna1}
J.\ Bernabeu, G.\ A.\ Gonzalez--Sprinberg, M.\ Tung and  J.\ Vidal,
Nucl.\ Phys.\ {\bf B436} (1995) 474. 

\bibitem{reg:dim} G.\ 't Hooft and M.\ Veltman, Nucl.\ Phys.\
{\bf B44} (1972) 189; C.\ G.\ Bollini and J.\ J.\ Giambiagi,
Nuovo Cim.\ {\bf 12B} (1972) 20.

\bibitem{feyn} R.\ Mertig, M.\ Bohm and A.\ Denner, Comput.\ Phys.\
Commun.\ {\bf 64} (1991) 345.

\bibitem{zep} K.\ Hagiwara, S.\ Ishihara, R.\ Szalapski and D.\
Zeppenfeld, Phys.\ Lett.\ {\bf B283} (1992)  353; Phys.\ Rev.\
D{\bf 48} (1993) 2182.

\bibitem{g-2exp} J. Bailey {\it et al.}, Nucl.\ Phys.\ {\bf B150} (1979) 1;
E.\ R.\ Cohen and  B.\ N.\ Taylor,   Rev.\ Mod.\ Phys.\ {\bf 59} (1987) 1121.

\bibitem{g-2QED} T.\ Kinoshita, B.\ Ni{\u{z}}i\'c, and Y.\
Okamoto, Phys. Rev. Lett. {\bf 52} (1984) 717. For a review of
the QED calculations see: T.\ Kinoshita and W.\ J.\ Marciano,
``Quantum Electrodynamics'', T.\ Kinoshita (ed.), World
Scientific, Singapore (1990), p.\ 419 and references therein.

\bibitem{g-2EW}
G.\  Altarelli, N.\  Cabibbo, and L.\  Maiani, Phys. Lett. {\bf B40} (1972)
415; I.\ Bars and M.\ Yoshimura, Phys. Rev. {\bf D6} (1972) 374;
K.\ Fujikawa, B.\ W.\  Lee and A.\ I.\ Sanda, Phys. Rev. {\bf D6} (1972)
2923; R.\ Jackiw and S.\ Weinberg, Phys. Rev. {\bf D5} (1972) 2473;
W.\ A.\  Bardeen,  R.\ Gastmans and B.\ E.\ Lautrup,  Nucl. Phys. {\bf B46} 
(1972) 319; T.\ V.\  Kukhto, E.\ A.\  Kuraev, A.\ Schiller and
Z.\ K.\ Silagadze, Nucl. Phys. {\bf B371} (1992) 567;
A.\ Czarnecki, B.\ Krause and W.\ J.\ Marciano, Phys. Rev. {\bf D52} (1995) 
2619 and Phys. Rev. Lett. {\bf 76} (1996) 3267;
S.\  Peris, M.\  Perrottet and E.\ de Rafael, Phys. Lett. {\bf B355} (1995) 523.

\bibitem{g-2had}
J.\ Calmet, S.\ Narison, M.\ Perrottet and E.\ de Rafael,
Phys. Lett. {\bf B61} (1976) 283; Rev. Mod. Phys. {\bf 49} (1977) 21; 
T.\ Kinoshita, B.\ Ni{\u{z}}i\'c and Y.\ Okamoto,
Phys. Rev. {\bf D31} (1985) 2108;
E.\ de Rafael, Phys. Lett. {\bf B322} (1994) 239;
E.\ Pallante, Phys. Lett. {\bf B341} (1994) 221;
M.\ Hayakawa, T.\ Kinoshita and  A.\ I.\ Sanda,
Phys. Rev. Lett. {\bf 75} (1995) 790;
J.\ Bijnens, E.\ Pallante and J. Prades,
Phys. Rev. Lett. {\bf 75} (1995) 1447; 
S.\ Eidelman and F.\ Jegerlehner, Z. Phys. {\bf C67} (1995) 585;
K.\ Adel and F.\ J.\ Yndur\'ain, Univ. Aut\'onoma de Madrid
preprint, FTUAM 95--32 , hep--ph/9509378 (1995).

\bibitem{futu}
B.\ L.\ Roberts, Z.\ Phys.\ C56 (1992) S101.

\end{references}
\end{document}